\documentclass[journal]{IEEEtran}
\usepackage{cite}
\usepackage{amsmath,amssymb,amsfonts}
\usepackage{algorithmic}
\usepackage{graphicx}
\usepackage{hyperref}

\usepackage{multirow}
\usepackage{booktabs,siunitx}
\usepackage{makecell}

\usepackage{balance}
\usepackage{flushend}

\usepackage{graphicx}

\usepackage{url}
\usepackage{hyperref}

\begin{document}
\title{Details Preserving Deep Collaborative Filtering-Based Method for Image Denoising}
\author{
	\textsuperscript{1}Basit~O. Alawode, \textsuperscript{2}Mudassir Masood, \textsuperscript{3}Tarig Ballal, and \textsuperscript{4}Tareq Al-Naffouri.\\
	\textsuperscript{1,2}King Fahd University of Petroleum and Minerals, Dhahran, Saudi Arabia. \\
	\textsuperscript{3,4}King Abdullah University of Science and Technology, Thuwal, Saudi Arabia. \\
	Email: {\{\textsuperscript{1}g201707310, \textsuperscript{2}mudassir\}@kfupm.edu.sa, 
		\{\textsuperscript{3}tarig.ahmed, \textsuperscript{4}tariq.alnaffouri\}@kaust.edu.sa}
}
\maketitle

\begin{abstract} 
In spite of the improvements achieved by the several denoising algorithms over the years, many of them still fail at preserving the fine details of the image after denoising. This is as a result of the smooth out effect they have on the images. Most neural network-based algorithms have achieved better quantitative performance than the classical denoising algorithms. However, they also suffer from the qualitative (visual) performance as a result of the smooth out effect. In this paper, we propose an algorithm to address this shortcoming. We propose a deep collaborative filtering-based (Deep-CoFiB) algorithm for image denoising. This algorithm perform collaborative denoising of image patches in the sparse domain using a set of optimized neural network models. This result in a fast algorithm that is able to excellently obtain a trade off between noise removal and details preservation. Extensive experiments show that the Deep-CoFiB performed quantitatively (in terms of PSNR and SSIM) and qualitatively (visually) better than many of the state-of-the-art denoising algorithms.
\end{abstract}


\section{INTRODUCTION}
\label{sec:intro}

The advancements in technology has made it possible to obtain crystal clear images with image capturing equipment. However, these equipment have very little control over the environment in which such images are taken. For instance, an image taken in a foggy environment. The imaging equipment could also introduce noise into the image. This implies that noise in images cannot be totally removed even with technologically advanced equipments. However, application area such as medical, astronomy, etc. require the use of noise free images. The process of removing noise in images is termed image denoising. Despite the long active efforts of researchers to develop several image denoising algorithms which have resulted in gradual improvement in performance, image denoising still remains an active research area till today. This is due to the need to create noise free images for several practical applications which imaging equipments have no panacea to automatically remove.

A noisy image $ \mathbf{Y} $ is a combination of the original image $ \mathbf{X} $ and a noise component $ \mathbf{W} $. This combination is considered to be additive as shown equation (\ref{eq:noisyImage}). To model $ \mathbf{W} $, its random and independent entries are drawn from the same probability distribution, e.g. Gaussian. As such, $ \mathbf{W} $ is typically said to be an additive, white and Gaussian noise (AWGN).

\begin{equation}
\mathbf{Y = X + W.}
\label{eq:noisyImage}
\end{equation}

Over the years, several denoising algorithms have been proposed. Many took advantage of mathematical analytic tools and background knowledge of images to perform denoising. These set of algorithms are termed classical denoising algorithms. Denoising algorithms such as the non-local means (NLM) \cite{buades2005}, block-matching 3D transform (BM3D) \cite{Dabov2007a}, K-singular value decomposition (K-SVD) \cite{Aharon2006a}, the collaborative support agnostic recovery (CSAR)\cite{Behzad2017}, non-local sparse model \cite{mairal2009}, etc. belong to this category. Many of these algorithms make use of image patches to perform denoising, except the NLM algorithm which uses similarities in the noisy image's pixels values to perform denoising. An image patch is a small part of the original image. A patch (an array of pixel values) usually contains more information about the image than a pixel (a single) value. This is why patch-based algorithms tend to perform better than pixel-based counterparts. Of note in the classical denoising algorithms is the BM3D \cite{Dabov2007a}. It denoises an image by transforming patches into a 3-D wavelet (sparse) domain and performing hard-thresholding on the sparse components of the noisy patches. Due to its effectiveness, it was used as a denoising benchmark for a very long time.

Recently, deep learning (DL) has recorded tremendous success in areas such as computer vision, natural language processing and forecasting, etc \cite{William2018}. This has propelled researchers to apply DL techniques to image denoising. At the heart of DL is the neural networks (NNs). As such, image denoising algorithms using DL are also known as NN-based denoising algorithms. This has resulted in several algorithms such as the fast feed-forward NN (FFDNet) \cite{Zhang2018}, deep convolutional NN (DnCNN) \cite{Zhang2017}, trainable reaction diffusion (TRND) \cite{chen2017}, block-matching convolutional NN (BMCNN) \cite{ahn2018}, etc. One notable algorithm is the DnCNN. It takes advantage of advanced DL techniques such as Residual learning \cite{He2016} and batch normalization \cite{Szegedy2015a} with several layers of NN convolution to perform denoising. The use of DL in denoising has seen the performance of the state-of-the-art classical denoising algorithms surpassed. This has put the denoising benchmark power on the NN-based algorithms.

Despite the improved performance of the NN-based algorithms, NNs are sometimes referred to as black boxes. This makes it analytically difficult for researchers to explain why such algorithms performed well. However, in a bid to offer some explanations and to get more performance out of the classical algorithms, researchers have tried to incorporate some elements of NN into the denoising pipelines of the classical denoising algorithms. This has resulted in neuro-classical denoising algorithms such as the BM3D-Net \cite{Yang2018a}, and the Deep K-SVD \cite{Scetbon2019a} denoising algorithms. The BM3D-Net transforms the classical BM3D algorithm into a set of learnable convolution layers while the Deep K-SVD used a set of multilayer perceptrons (MLPs) to learn the value of some parameters used in the classical K-SVD denoising algorithm. These algorithms achieved better performance than their classical counterparts.

We had earlier developed a collaborative filtering-based (CoFIB) image denoising algorithm \cite{Basit2020b}. The CoFiB algorithm is a patch-based classical denoising algorithm. It takes advantage of the fact that similar patches have similar components in the sparse domain to perform a weighted sparse domain collaborative denoising. This resulted in an algorithm that was able to address one of the shortcomings of denoising algorithms - inability to preserve image details. The CoFiB algorithm was also able to perform well for images of any given resolution. However, this performance comes with an increased denoising time relative to most of the other algorithms. Relying on the improved and faster performance of the NN-based algorithms, in this paper, we propose to extend the denoising performance of the CoFiB algorithm by translating its denoising pipeline into a NN-based version. This algorithm we tagged as the Deep-CoFiB denoising algorithm. Our contribution is to leverage on NNs to obtain a fast algorithm capable of preserving details in images after denoising by rethinking the CoFiB algorithm in the NN manner.

The rest of this paper is organized as follows. Section \ref{sec:cofib} provides the details of the CoFiB denoising algorithm. In Section \ref{sec:dcofib}, we provide the details of the proposed Deep-CoFiB algorithm. Detail experimental results are presented in Section \ref{sec:exp} and finally, we conclude in Section \ref{sec:con}.

\section{THE COFIB DENOISING ALGORITHM}
\label{sec:cofib}

\begin{figure*}
	\centering
	\includegraphics[width=\textwidth]{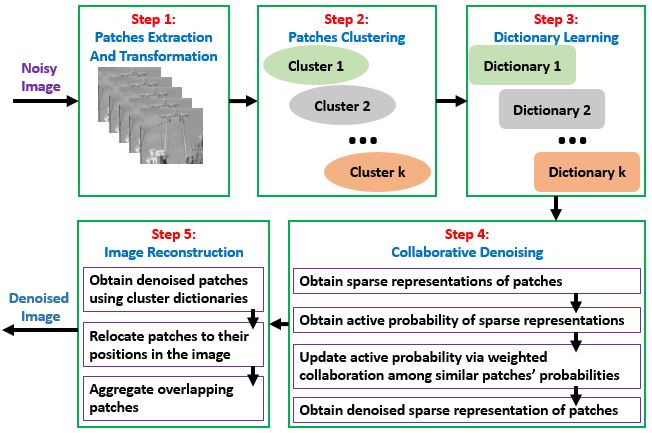}
	\caption{Flowchart of the CoFiB Algorithm.}
	\label{fig:cofibFlowchart}
\end{figure*}

The CoFiB \cite{Basit2020b} algorithm performs collaborative denoising in 5 steps. This is as shown in Fig. \ref{fig:cofibFlowchart}. The algorithm performs denoising of patches in the sparse domain. This is done by allowing the sparse representations of similar patches to collaborate in a weighted approach to denoise a given patch. Once all patches in the image have been denoised, they are replaced to their original positions to produce the denoised image. We briefly describe these steps below. 

\begin{enumerate}
	\item \textbf{Patches Extraction and Transformation:} Overlapping patches are extracted from the image. These patches are then transformed to an intensity invariant set by dividing each patch by its maximum absolute pixel value.
	
	\item \textbf{Patches Clustering:} As the algorithm collaborates among similar patches' sparse representations in the sparse domain, similar patches are clustered together in order to have similar patches better represented in the sparse domain. This is done using the k-means clustering algorithm \cite{jain2010}.
	
	\item \textbf{Dictionary Learning:} For each cluster, an overcomplete dictionary is learned. This is done using the K-SVD dictionary learning algorithm \cite{Aharon2006a}.
	
	\item \textbf{Collaborative Denoising:} Given the cluster dictionaries, the sparse representations and active location probabilities of the patches are obtained using the support agnostic Bayesian matching pursuit (SABMP) sparse recovery algorithm \cite{Masood2013a}. A patch is denoised by updating/denoising its sparse representation's active location probabilities. This is done by performing a weighted collaboration among the active location probabilities of similar patches' sparse representations. The updated/denoised active probability of the patch is then fed back to the SABMP algorithm to obtain the denoised patch's sparse representation.
	
	\item \textbf{Image Reconstruction:} The denoised sparse representations of all the patches are transformed back to the spatial domain using the respective cluster dictionary of each patch to obtain the denoised patches. In order to reconstruct the denoised image, the denoised patches are simply replaced back to their original locations in the noisy image. Noting that these patches overlap, they are then aggregated to obtain the denoised image.  
\end{enumerate}

To leverage on the faster and better quantitative performance of the NN-based algorithms, we propose to translate the CoFiB denoising pipeline into a NN-based version. The details of the proposed algorithm is as presented in the Section below.

\section{THE DEEP-COFIB DENOISING ALGORITHM}
\label{sec:dcofib}

\begin{figure*}
	\centering
	\includegraphics[width=\textwidth]{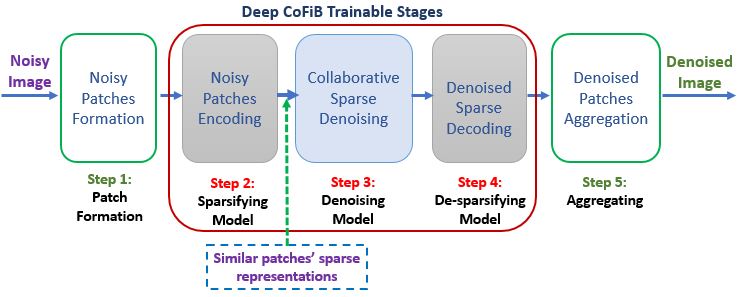}
	\caption{Flowchart of the Proposed Deep-CoFiB Algorithm.}
	\label{fig:dcofibFlowchart}
\end{figure*}

Basically, the CoFiB denoising pipeline can be broken down as follows - 1. Overlapping patches formation, 2. Transformation of the patches from the spatial domain into the sparse domain, 3. Collaboration among similar patches in the sparse domain to perform denoising, 4. Transformation of the denoised sparse representations back to the spatial domain, and 5. Replacing patches back to their original position and averaging overlapping patches. This breakdowns form the flow of our proposed Deep-CoFiB denoising algorithm. This is as shown in Fig. \ref{fig:dcofibFlowchart}. We now proceed to describe these steps as follows.   

\subsection{Step 1: Patches Formation}

Similar to the CoFiB algorithm, given a noisy image of size $ N \times M $, overlapping patches of size $ n \times n $ each are obtained. These patches are then vectorized such that each patch is of size $ n^{2} \times 1 $.

\subsection{Step 2: Noisy Patches Encoding}

We transform the vectorized patches from the spatial domain to the sparse domain. This was done using fully connected NN layers resulting in a model we termed the sparsifier. It takes in a vectorized noisy patch of size $ n^{2} \times 1 $ and outputs the encoded sparse representation of the patch. This has a dimension of $ m \times 1 $.

\subsection{Step 3: Collaborative Sparse Domain Denoising}

\begin{figure*}
	\centering
	\includegraphics[width=\textwidth]{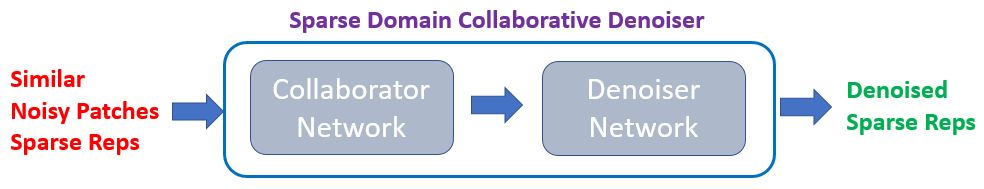}
	\caption{The Collaborative Sparse Domain Denoiser Block.}
	\label{fig:collabDenoiser}
\end{figure*}

Every noisy patch obtained from the image is denoised using the collaborative sparse domain denoiser. This is done as follows. For each noisy reference patch, $ d-1 $ similar noisy patches are obtained. Patches similarity is obtained using any suitable distance measure such as the Euclidean distance. To search for similarity to the reference patch, a local window $ S \times S $ around the reference patch is used. This is done to reduce the computation complexity of finding similar patches across the whole image. The reference patch and the similar patches are then fed into the sparsifier model (in step 2 above) to obtain their sparse representations. These sparse representations are then fed into the sparse domain collaborative denoiser to obtain the denoised sparse representation of the reference patch. This is as shown in Fig. \ref{fig:collabDenoiser}.

As shown in Fig. \ref{fig:collabDenoiser}, the collaborative sparse domain denoiser consist of two inner networks - 1. the collaborator network, and 2. the denoiser network. We now proceed to describe each of them below.

\begin{enumerate}
	\item Collaborator Network: This is a custom trainable element-wise collaborator network. This network has the same size as the similar sparse representations used for collaboration. In eq. \ref{eq:trainablecollaboration}, we show how the element-wise trainable collaboration is performed. To collaborate among the sparse representations of the $ d $ similar patches (the reference patch and the $ d-1 $ similar patches), we concatenated their sparse representations into a size of $ m \times d $ sparse matrix. This is then fed into the trained collaborator network to obtain the collaborated sparse representation of size $ m \times 1 $. 
	
	\begin{equation}
		\begin{aligned}
			\textrm{Similar noisy sparse reps}, \quad X = m \times d \\
			\textrm{Trainable collaboration weights}, \quad C = m \times d \\
			\textrm{Element-wise collaboration} \quad = C \bigodot X = m \times d \\
			\textrm{Collaborated sparse rep (average)} \quad = m \times 1
		\end{aligned}
		\label{eq:trainablecollaboration}
	\end{equation}
	
	\item Denoiser Network: The denoiser network consists of fully connected NN layers. The collaborated sparse representation is then fed into the denoiser network to obtain the denoised sparse representation of the reference patch. 
	
\end{enumerate}

The combination of these networks (the collaborator and the denoiser networks) is what we referred to as the denoiser model in Fig. \ref{fig:dcofibFlowchart}, step 3. The denoiser model therefore takes the sparse representation of a reference noisy patch and the similar patches to produce the denoised sparse representation of the reference patch. 

\subsection{Step 4: Denoised Sparse Decoding}

In this step, the denoised sparse representation is transformed from the sparse domain back into the spatial domain to obtain the denoised vectorized patch. This is also achieved using fully connected NN layers resulting in a model we termed the desparsifier. It takes in the denoised sparse representation of size $ m \times 1 $ and outputs the denoised vectorized patch of size $ n^{2} \times 1 $. This is then reshaped into a dimension of $ n \times n $ to obtain the denoisd patch.

\subsection{Step 5: Denoised Patches Aggregation}

All patches obtained from the noisy image are denoised using the trained models of steps 2, 3, and 4 above. The denoised patches are then replaced back to their original positions in the image. All overlapping patches are then averaged to obtain the denoised image.

\section{EXPERIMENTAL RESULTS}
\label{sec:exp}

We performed extensive experiments on the proposed algorithm. We compared the performance with the classical techniques such as NLM \cite{buades2005}, BM3D \cite{Dabov2007a}, K-SVD \cite{Aharon2006a}, and the CoFiB \cite{Basit2020b}. We also compared with the NN-based techniques tsuch as the FFDNet \cite{Zhang2018} and the DnCNN \cite{Zhang2017} algorithms. We give details of our experimental settings in the following section.

\subsection{Experimental Settings}

\begin{enumerate}
	\item Design Parameters: The patch size used in this research is $ 5 \times 5 $. This implies $ n = 5 $. In sparse coding, in order to ably reconstruct the patches from its' sparse code, $ m \geqq n^{2} $ and $ m \gg 2 \times \textrm{sparsity rate} $ \cite{Stankovic2017}. Since this is a NN approach, we found it practically challenging to specify a precise sparsity rate. However, our extensive research provided an average sparsity rate of 30 given $ m = 100 $. This satisfies the above sparsity condition. To find similar patches, local window size of $ S \times S = 50 \times 50 $ and the number of patches for collaboration $ d = 5 $ were used.
	
	\item Training Data: For generating the training patches, we used the training images \cite{dncnntraining} that were also used for training the DnCNN denoising algorithm \cite{Zhang2017}. The training images are a set $ 100 $ images randomly selected from a set of $ 400 $ clean gray scale images of size $ 100 \times 100 $. For each clean image, we generated vectorized patches of size $ n^{2} \times 1 $ each. As such, each image produced $ 100 \times 100 = 10,000 $ clean patches. The 100 clean images, therefore, generated $ 1,000,000 $ clean patches. 
	
	To generate the noisy images, we added white Gaussian noise (WGN) of known variance (for example $ \sigma = 25 $) to the clean patches. As such, we trained using a set of $ 1,000,000 $ clean patches and $ 1,000,000 $ noisy patches.
	
	\item The Trainable Models: As shown in Fig. \ref{fig:dcofibFlowchart} and described earlier, the Deep-CoFiB has only three trainable models - 1) the sparsifier, 2) the denoiser, and 3) the desparsifier models. Our aim is to keep these models as simple as possible without compromising the denoising performance of the algorithm. As such, the description of the models are as follows.
	
	\begin{enumerate}
		\item The sparsifier model has only 2 hidden fully connected NN layers. The input layer has $ n^{2} = 25 $ nodes. Each of the hidden and output layers has $ m = 100 $ nodes. Each of the nodes in the hidden and output layers has a bias. These correspond to $ 22,800 $ trainable parameters for the sparsifier model. 
		
		\item The denoiser model consists of the collaborator and the denoiser networks. The collaborator network has $ d \times m = 500 $ parameters. The denoiser network consists of 10 fully connected NN layers. Each layer has $ m = 100 $ nodes with each node having its own bias. As such, the denoiser network has $ 101,000 $ parameters. The total number of parameters for the denoiser model is therefore $ 101,500 $.
		
		\item Similar to the sparsifier model, the desparsifier model also has only 2 hidden fully connected NN layers. Each of the input layer and the hidden layers has $ m = 100 $ nodes. The output layer has $ n^{2} = 25 $ nodes. Each of the nodes in the hidden and output layers also has a bias. These correspond to $ 22,725 $ trainable parameters for the desparsifier model. The proposed Deep-CoFiB algorithm, therefore, has a total of $ 147,025 $ trainable parameters.
	\end{enumerate}

	\item Training: NN performs best with spatially connected data \cite{sparseQuora} making it challenging for them to find connections among sparsely connected data. In our algorithm, we will be dealing with sparse data. As such, to train the three models, we employed a joint training strategy to deal with this known NN challenge. This was achieved in two stages as described below.
	
	\begin{enumerate}
		\item The sparsifier and desparsifier models are combined resulting in a joint model 1 that takes a noisy patch and outputs a clean patch. This model 1 is then trained using the noisy data at the input and the corresponding clean data at the output. 
		
		\item To train the denoiser model, we combined the denoiser and the already trained desparsifier model. This resulted in a joint model 2 that takes in sparse representations of noisy patches and outputs a denoised patch. The trained desparsifier model is obtained from the joint model 1. During training, the weights of the already trained desparsifier model was frozen. This ensures that only the denoiser model was being trained. For training, we obtain the sparse representations of a reference noisy patch and its similar patches using the trained sparsifier and fed it into the joint model 2. This is then trained against the corresponding reference clean patch.
	\end{enumerate}

	For both training stages, we specified a training batch size of 100 and trained the joint networks for only 200 epochs. The mean squared error (MSE) was used as the loss function. For weights update, the Adam optimizer \cite{kingma2015} was employed. The training time and how it compared with other NN-based algorithm is discussed in Section \ref{sec:trainingTime}.
	
	\item Code Development: We used Python deep learning application programming interface (API) Keras library \cite{keras} with TensorFlow backend to develop the proposed Deep-CoFiB  models. The training was performed using 2 Nvidia GTX 1080 Ti GPUs running on a Linux-based computing cluster with a preallocated 32GB of RAM. Testing was performed on a PC with Intel(R) Core™ i7-7500U CPU running at 2.70GHz with 16GB of RAM.  
	
\end{enumerate}

\subsection{Performance Comparison on Different Noise Variances}
\label{sec:cofibresultsdiffsnr}

We compared the proposed Deep-CoFiB algorithm with other algorithms on different noise levels (variances). We used one of the commonly used images in literature for this experiment - the boat image of size $ 256 \times 256 $. WGN was added to the original image. Table \ref{tab:dcofibdiffsnrTable} shows a comparison of the performance of the different algorithms on different noise variances. 

\begin{figure*}
	\centering
	\includegraphics[width=0.7\textwidth, height=0.3\textheight]{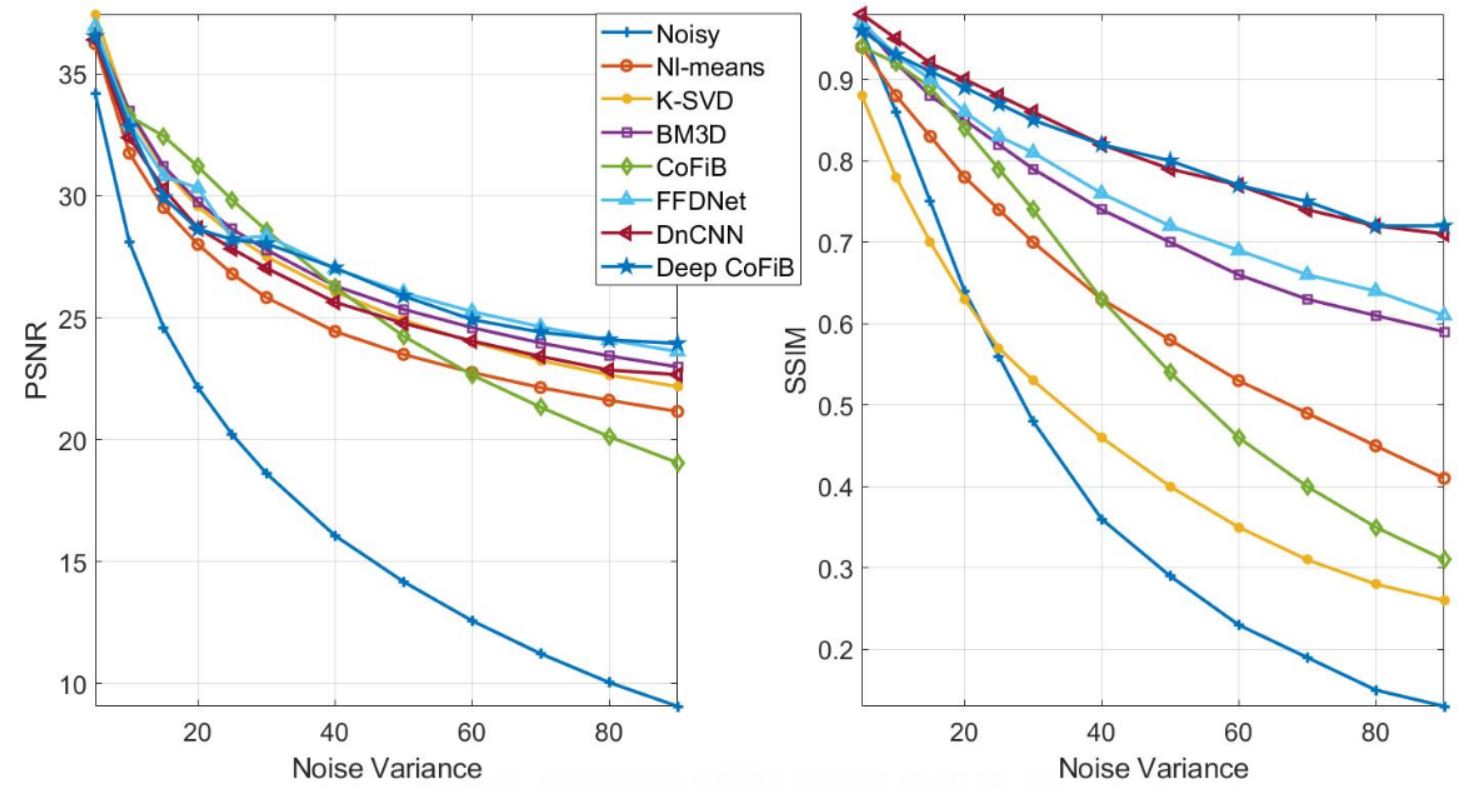}
	\caption{Graphical: Performance Comparison for Different Noise Variances Using the Boat Image ($ 256 \times 256 $)}
	\label{fig:dcofibdiffsnrgraph}
\end{figure*}

It can be observed from the table that the performance of the NN-based algorithms are generally better than the classical counterparts except for the CoFiB algorithm in few cases. Our proposed Deep-CoFiB algorithm can be observed to perform comparably with the other NN-based algorithms at lower noise level. At higher noise level, it performed better than all the algorithms. This is as a result of the detail preserving nature of the algorithm. These results are presented graphically as shown in Fig. \ref{fig:dcofibdiffsnrgraph}.

We visually compare the results for 3 noise levels (low, medium, and high noise levels). Here, only the BM3D, CoFiB, DnCNN, and  Deep-CoFiB results are shown. In Table \ref{tab:dcofibdiffsnrVar10}, we visually show the results of these algorithms for a low noise variance of $ \sigma = 10 $. Similarly, we showed for an average noise variance of $ \sigma = 40 $ in Table \ref{tab:dcofibdiffsnrVar40} and Table \ref{tab:dcofibdiffsnrVar90} showed that of severe noise level of variance $ \sigma = 90 $. In all these tables, we can notice that the CoFiB and the proposed Deep-CoFiB algorithms preserved the details of the image. This is indeed apparent in the presence of severe noise ($ \sigma = 90 $).

\subsection{Performance Comparison of Different Algorithms on Different Image Resolutions}
\label{sec:cofibresultsdiffresolution}

In this section, we compare the performance of the proposed Deep-CoFiB algorithm with other denoising algorithms on different image resolutions. We generated images of three different resolutions of $ 64\times64 $, $ 128\times128 $, and $ 256\times256 $. WGN was added to these images with SNR of 20dB and then denoised using the different algorithms. Table \ref{tab:dcofibdiffResTable} shows the results of the various algorithms on the different boat image resolutions. It can be observed that the proposed algorithm performed well for all the image resolutions considered. More particularly, it achieved significantly better PSNR and SSIM for images of lower resolutions beating the DnCNN in the process. 

In addition, we visually compared the results as shown in Table X which can be found here \cite{dcofib_more_result}. From this table, it can clearly be seen that for lower resolution images of $ 64\times64 $ and $ 128\times128 $, the denoised output was completely distorted by the other algorithms, including the DnCNN. Similar to the CoFiB, the proposed Deep-CoFiB was equally able to preserve the details of the images (i.e. lines, the edges, flat regions, etc.). Particularly, if we compare DnCNN with the Deep-CoFiB, we realized that the DnCNN generated smoothened out results whereas the Deep-CoFiB gave clearer and undistorted denoised images. Many algorithms seem not to perform well on images of lower resolutions as they tend to loose most of the image details.

\subsection{Performance Comparison of Different Algorithms on Different Images}
\label{sec:cofibresultsdiffimages}

We compared the proposed algorithm’s performance with other algorithms on different benchmark images. This is as shown in Table \ref{tab:dcofibdiffImagesTable}. Again, WGN corresponding to SNR of 20dB was added to all the images. From the table, we can notice that although, the DnCNN tend to perform quantitatively better than the proposed algorithm. However, qualitatively (visually), our proposed algorithm performed better. In Table XI, which can be found here \cite{dcofib_more_result}, we show the visual results of the boat, cameraman, and mandrill images. It can also be observed from this table that the Deep-CoFiB result presented a more realistic denoised version of the original image.

\subsection{Trainable Parameters Comparison}

The denoising architecture of the FFDNet and DnCNN are similar. The only difference in their architectures is the fact that DnCNN learns the noise in the image using residual learning, whereas the FFDNet learns the actual denoised image. As such, the total number of parameters for the two algorithms are the same as shown in Table \ref{tab:nnAlgoParameterCompare}. Due to the simple denoising architecture of our proposed Deep-CoFiB algorithm, it has a significantly reduced number of trainable parameters (67.11\% reduction) as compared to the other two high performing NN-based algorithms.

\begin{table}
	\centering
	\caption{NN-Based Algorithms Parameters Comparison}
	\begin{tabular}{l|c}
		\hline
		\textbf{Algorithm} & \textbf{Number of Parameters} 
		\\
		\hline
		& \\
		\textbf{DnCNN} & 447,057  \\
		& \\
		\textbf{FFDNet} & 447,057 \\
		& \\
		\textbf{Deep-CoFiB} & \textbf{147,025} \\
		\hline
		\hline
	\end{tabular}
	\label{tab:nnAlgoParameterCompare}
\end{table}

\subsection{Training Time Comparison} 
\label{sec:trainingTime}

The training time for the considered NN-based algorithms is as shown in Table \ref{tab:nnAlgoTrainTimeCompare}. Owing to the structural similarity of the DnCNN and the FFDNet, they have comparable training time. From this table, we can notice that our proposed Deep-CoFiB exhibit significantly lower training time as compared to the other two algorithms.

\begin{table}
	\centering
	\caption{NN-Based Algorithms Training Time Comparison}
	\begin{tabular}{l|c}
		\hline
		\textbf{Algorithm} & \textbf{Training Time (minutes)} 
		\\
		\hline
		& \\
		\textbf{DnCNN} & 230  \\
		& \\
		\textbf{FFDNet} & 250 \\
		& \\
		\textbf{Deep-CoFiB} & \textbf{145} \\
		\hline
		\hline
	\end{tabular}
	\label{tab:nnAlgoTrainTimeCompare}
\end{table}
	
\subsection{Denoising Time Comparison}

We also compared the denoising time of the different algorithms. This is as shown in Table \ref{tab:denTimeCompare}. The results shown is the average denoising time for 5 different images of size $ 512 \times 512 $ each. From the table, we can notice that CoFiB has a very high denoising time. This is because the algorithm needs to estimate and re-estimate noisy and denoised sparse representations using the SABMP algorithm. In addition, the algorithm needs to obtain trained dictionary using the K-SVD every time it is performing denoising. This is one of the main reason for the development of the Deep-CoFiB. The other is the limitations of the trained dictionary used. For NN-based algorithms, once the model has been trained, denoising becomes faster and consumes less computational resources. This is evident from the Table \ref{tab:denTimeCompare} where NN-based algorithm have comparably lesser denoising time.

\begin{table}
	\centering
	\caption{Denoising Time Comparison}
	\begin{tabular}{l|c}
		\hline
		\textbf{Algorithm} & \textbf{Denoising Time (minutes)} 
		\\
		\hline
		& \\
		\textbf{NLM} & 2.5  \\
		& \\
		\textbf{K-SVD} & 7  \\
		& \\
		\textbf{BM3D} & 5  \\
		& \\
		\textbf{CoFiB} & 40  \\
		& \\
		\textbf{FFDNet} & 1.68  \\
		& \\
		\textbf{DnCNN} & 2 \\
		& \\
		\textbf{Deep-CoFiB} & 5.5 \\
		\hline
		\hline
	\end{tabular}
	\label{tab:denTimeCompare}
\end{table}

\begin{table*}
	\centering
	\caption{Performance Comparison For Different Noise Variances Using the Boat Image ($ 256 \times 256 $)}
	\begin{tabular}{ c }
		\includegraphics[width=0.7\textwidth, height=0.3\textheight]{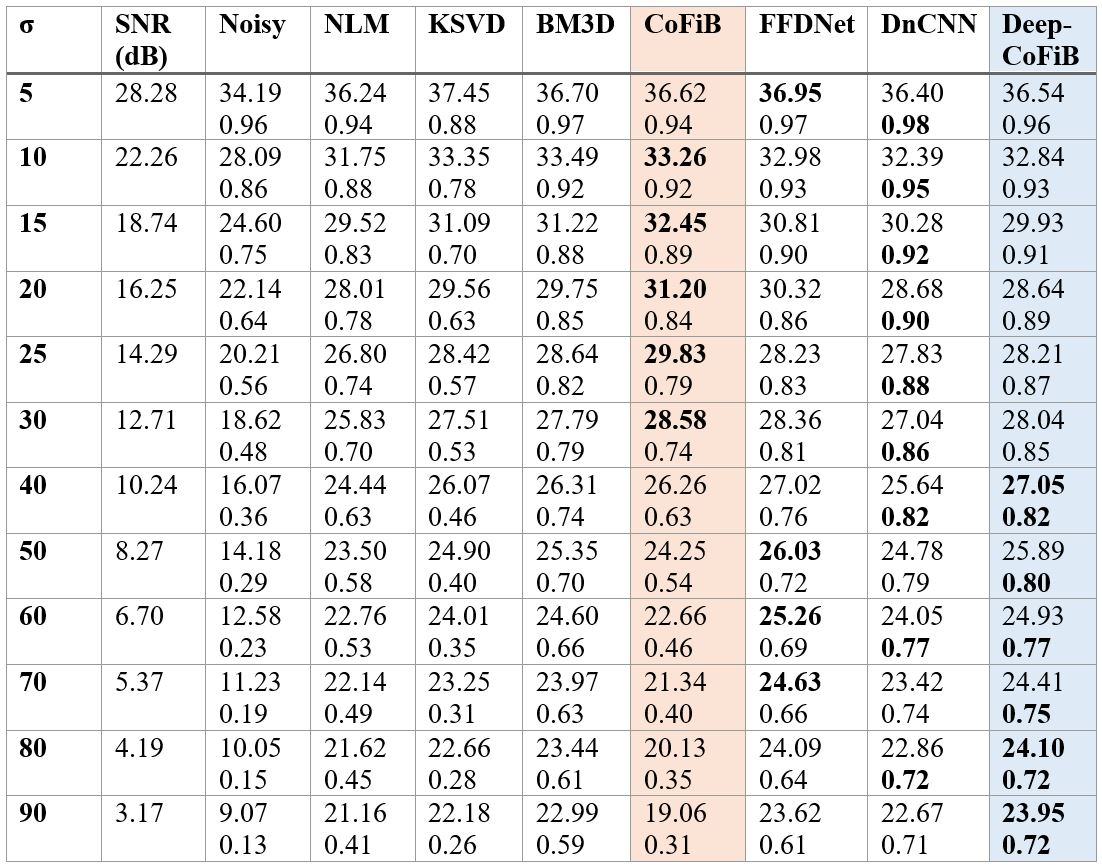}	
	\end{tabular}
	\label{tab:dcofibdiffsnrTable}
\end{table*}

\begin{table*}
	\caption{Visual Comparison of the Boat Image (Noise Variance = 10)} 
	\label{tab:dcofibdiffsnrVar10} 
	\begin{tabular}{c c c}
		\hline
		\hline
		(a) - Original & (b) - Noisy & (c) - BM3D \\
		\includegraphics[width= 0.3\textwidth]{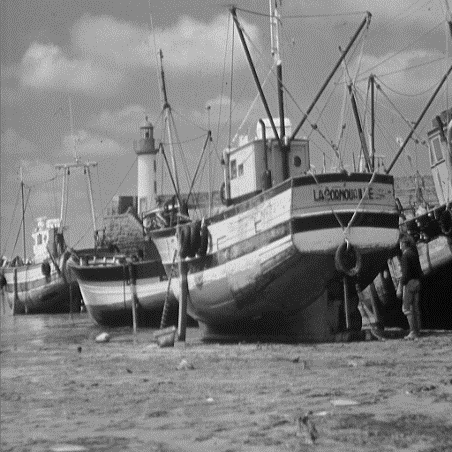} &	
		\includegraphics[width= 0.3\textwidth]{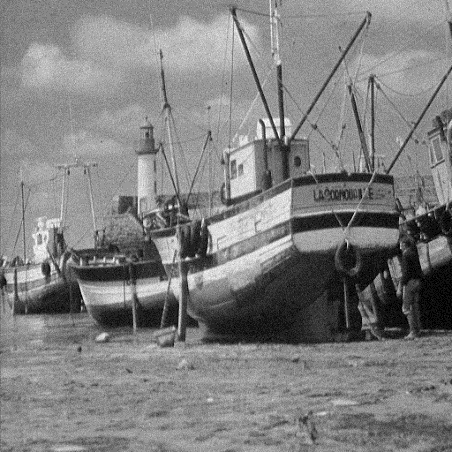}
		&
		\includegraphics[width= 0.3\textwidth]{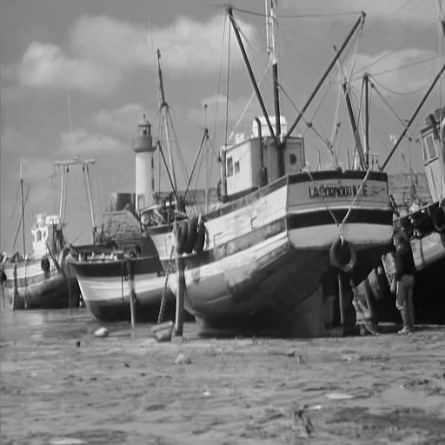}
		\\
		\hline
		(d) - CoFiB & (e) - DnCNN & (f) - Deep-CoFiB \\
		\includegraphics[width= 0.3\textwidth]{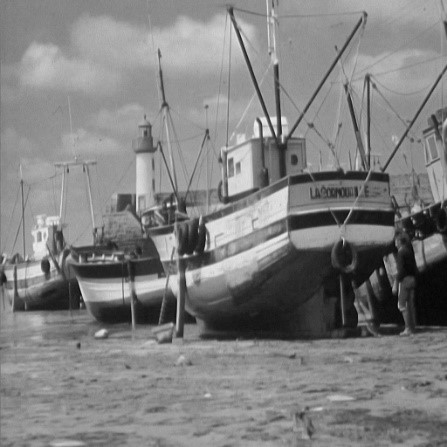} &	
		\includegraphics[width= 0.3\textwidth]{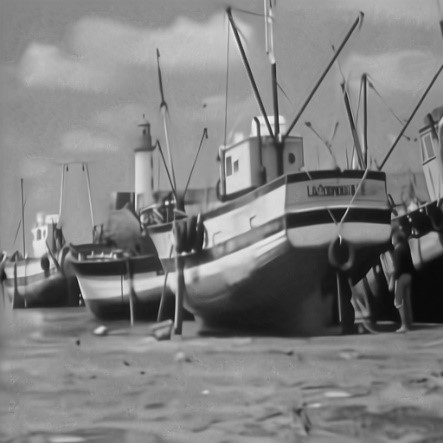}
		&
		\includegraphics[width= 0.3\textwidth]{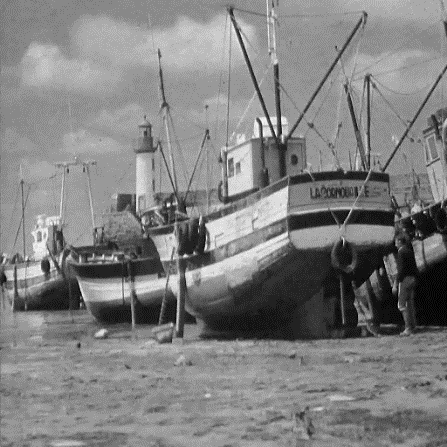}
		\\
		\hline
		\hline
	\end{tabular}
\end{table*}

\begin{table*}
	\caption{Visual Comparison of the Boat Image (Noise Variance = 40)} 
	\label{tab:dcofibdiffsnrVar40} 
	\begin{tabular}{c c c}
		\hline
		\hline
		(a) - Original & (b) - Noisy & (c) - BM3D \\
		\includegraphics[width= 0.3\textwidth]{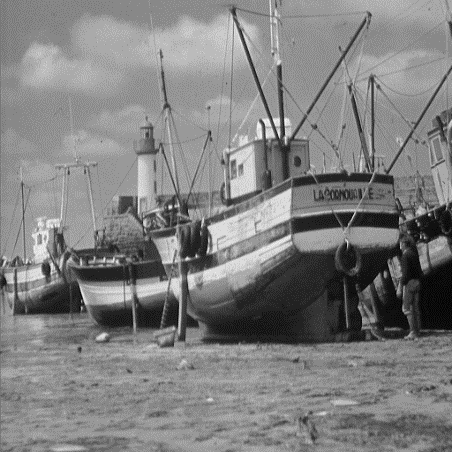} &	
		\includegraphics[width= 0.3\textwidth]{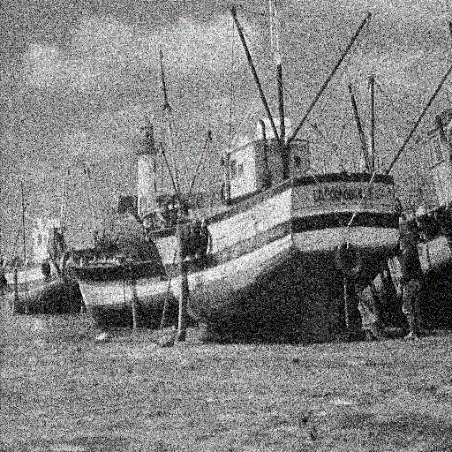}
		&
		\includegraphics[width= 0.3\textwidth]{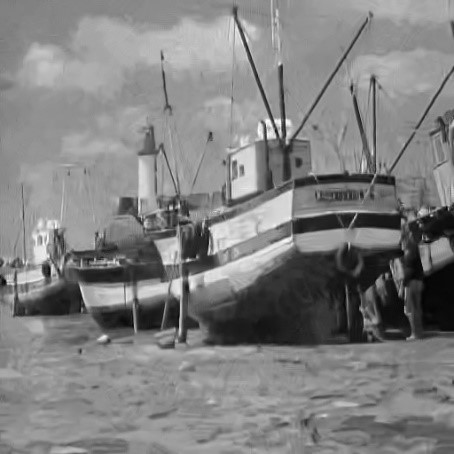}
		\\
		\hline
		(d) - CoFiB & (e) - DnCNN & (f) - Deep-CoFiB \\
		\includegraphics[width= 0.3\textwidth]{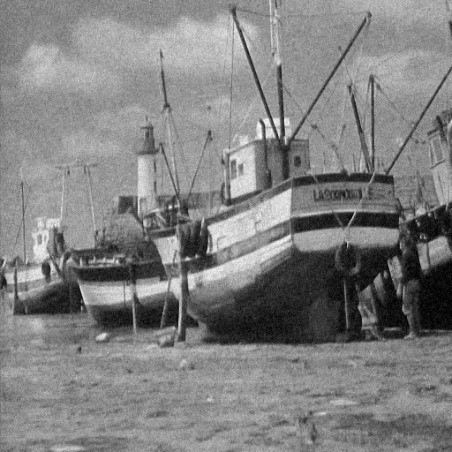} &	
		\includegraphics[width= 0.3\textwidth]{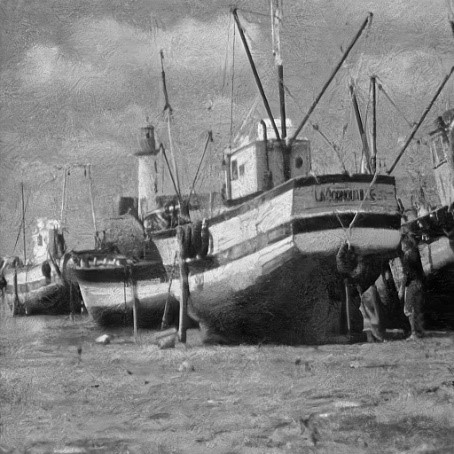}
		&
		\includegraphics[width= 0.3\textwidth]{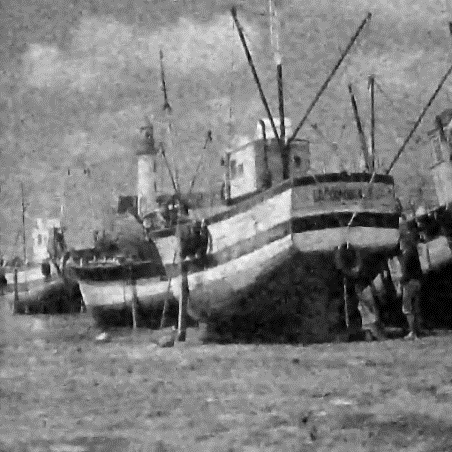}
		\\
		\hline
		\hline
	\end{tabular}
\end{table*}

\begin{table*}
	\caption{Visual Comparison of the Boat Image (Noise Variance = 90)} 
	\label{tab:dcofibdiffsnrVar90} 
	\begin{tabular}{c c c}
		\hline
		\hline
		(a) - Original & (b) - Noisy & (c) - BM3D \\
		\includegraphics[width= 0.3\textwidth]{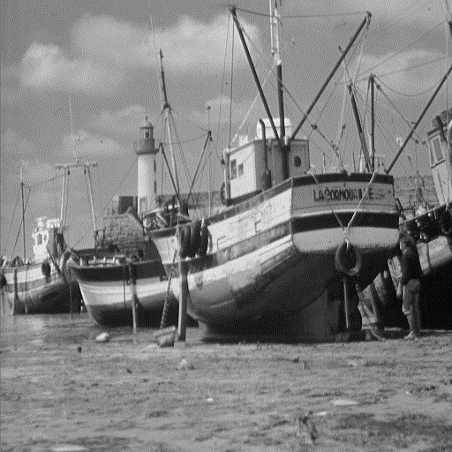} &	
		\includegraphics[width= 0.3\textwidth]{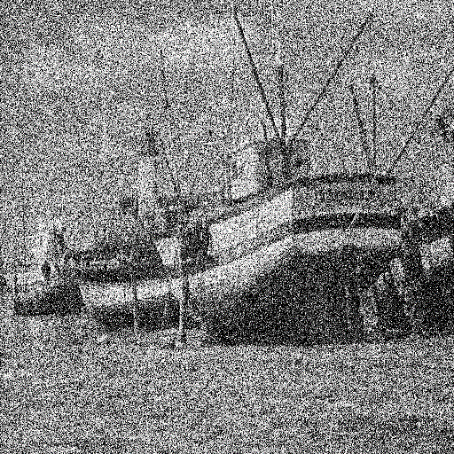}
		&
		\includegraphics[width= 0.3\textwidth]{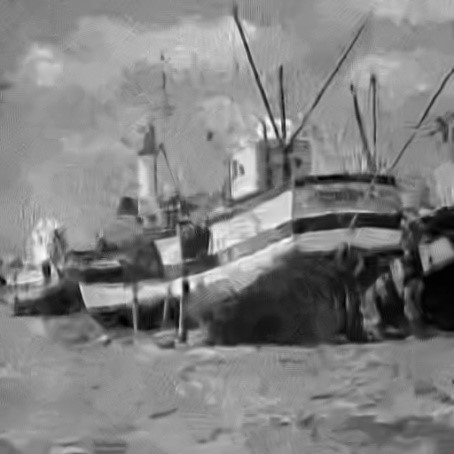}
		\\
		\hline
		(d) - CoFiB & (e) - DnCNN & (f) - Deep-CoFiB \\
		\includegraphics[width= 0.3\textwidth]{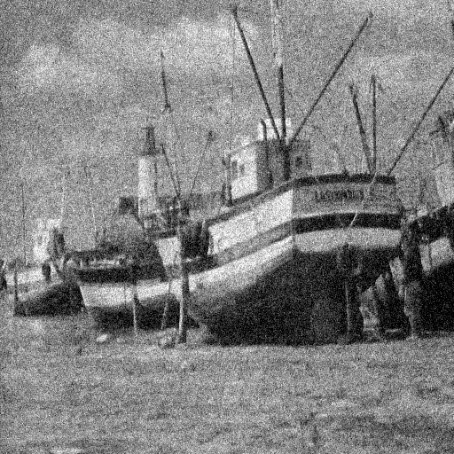} &	
		\includegraphics[width= 0.3\textwidth]{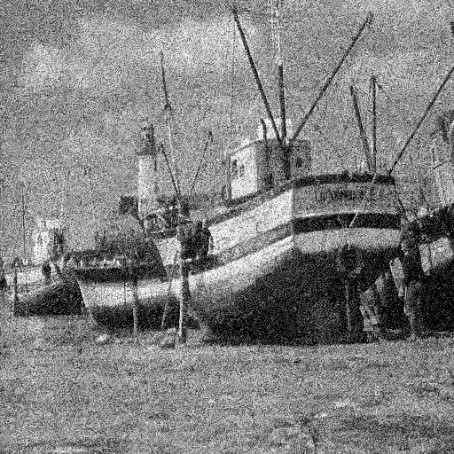}
		&
		\includegraphics[width= 0.3\textwidth]{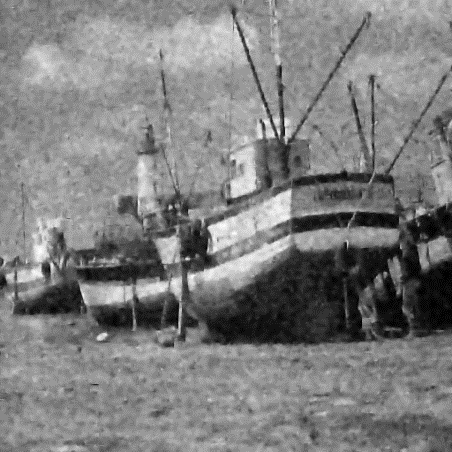}
		\\
		\hline
		\hline
	\end{tabular}
\end{table*}

\begin{table*}
	\centering
	\caption{Performance of Different Algorithms on Different Resolutions (Boat Image, SNR of 20dB)}
	\begin{tabular}{ c }
		\includegraphics[width= \textwidth]{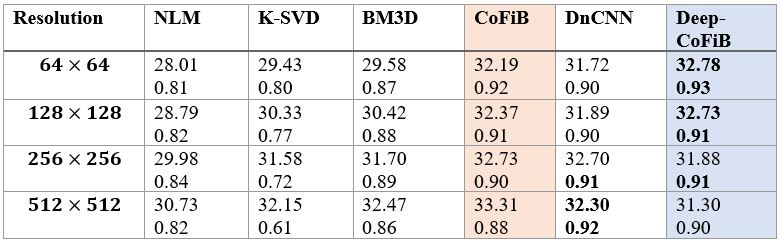}	
	\end{tabular}
	\label{tab:dcofibdiffResTable}
\end{table*}

\begin{table*}
	\centering
	\caption{Performance Comparison of Different Algorithms on Different Images (SNR = 20dB)}
	\begin{tabular}{ c }
		\includegraphics[width= \textwidth]{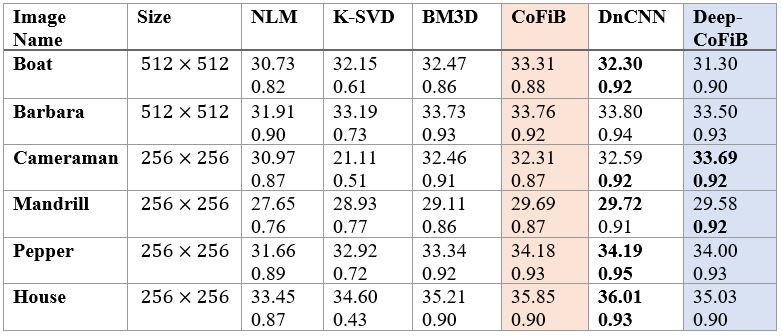}	
	\end{tabular}
	\label{tab:dcofibdiffImagesTable}
\end{table*}

\section{CONCLUSION}
\label{sec:con}

In this paper, we proposed a deep collaborative filtering-based (Deep-CoFiB) algorithm for image denoising. This algorithm is a direct translation of the collaborative filtering-based (CoFiB) denoising method into a neural network (NN)-based version. Similar to the CoFiB, the proposed Deep-CoFiB algorithm performs denoising in the sparse domain where similar patches' sparse representations are allowed to collaborate in a weighted approach. The proposed algorithm employed a trained set of NN models to sparsify and denoise the image patches. Several denoising algorithms tend to smooth out the denoised image which lead to loss of salient image components. The proposed Deep-CoFiB algorithm was able to preserve these image details as it was able to strike a balance between preserving the detail and complete noise removal. As an improvement to the CoFiB algorithm, it was able to quantitatively (in terms of PSNR and SSIM) surpass the CoFiB algorithm and performed comparably with other NN-based denoising algorithms. Furthermore, the algorithm was able to address the relatively high denoising time of the CoFiB algorithm. The proposed algorithm is a step in the right direction as researchers exploit ways to leverage on deep learning, due to its success in various fields, to improve the performance of classical denoising algorithms. 

\section{ACKNOWLEDGEMENT}
\label{sec:ack}

We gratefully acknowledge the support of the KAUST Supercomputing Lab for providing us with the computing cluster (ibex) and GPUs that were used to carry out this research.

\bibliographystyle{IEEEtran}
\bibliography{/Users/BASTECH-LPC/Documents/Mendeley/KAUST}

\end{document}